# A Minimal Bicomplex Extension of the Complex Scalar Algebra of Quantum Mechanics with an Ideal-Valued Sector

Ralf Otte
Faculty of Mechanical and Automotive Engineering
Institute for Automation Systems, Ulm University of Applied Sciences
Prittwitzstrasse 10, 89075 Ulm, Germany
ORCID: https://orcid.org/0009-0009-2293-8394
ralf.otte@thu.de

**Abstract**

Standard quantum mechanics takes the complex numbers as its scalar algebra. We ask whether this complex structure is algebraically fundamental or may instead form a distinguished embedded sector of a larger commutative and associative scalar algebra. Requiring the ordinary complex sector to remain intact while admitting an additional stable complex-like analytic sector turns the latter into a proper ideal, makes zero divisors unavoidable in finite dimension, and implies a lower bound of four real dimensions. The bicomplex algebra realizes this minimum, and under an additional compatibility assumption the four-dimensional realization is unique up to isomorphism.

The resulting algebra exhibits two distinct decompositions: the canonical decomposition into orthogonal idempotent ideals and a decomposition adapted to the embedded complex sector and the additional ideal-valued sector. The ordinary complex scalars therefore lie diagonally across the canonical idempotent decomposition. The ideal sector supports Taylor series, oscillatory exponentials, Fourier-type kernels, and Schrödinger-type expressions, and carries a formal ideal-valued eigenvalue structure. Restricting the Born rule to the embedded complex sector recovers its standard form. Whether this dual scalar architecture has physical significance beyond the algebraic framework remains an open question.

# 1 Introduction

Standard quantum mechanics is not merely written in the language of complex Hilbert spaces. Its linear superposition principle, phase structure, interference phenomena, Fourier representations, spectral theory, and Schrödinger dynamics are organized by the field of complex numbers. The choice of the complex scalar structure is therefore not a purely technical convention, but a structural assumption of the theory. This makes the scalar algebra itself a legitimate object of mathematical and physical analysis.

The empirical success of standard quantum mechanics is not questioned here. Precisely because the theory is successful, it is meaningful to ask whether the use of the complex numbers as scalar algebra is forced by the structural requirements of the theory, or whether it can be understood as the distinguished empirical sector of a larger compatible scalar algebra. The motivation is therefore not to correct standard quantum mechanics, but to examine the algebraic rigidity of one of its most basic structural assumptions: the use of the complex numbers as its scalar algebra. The aim is to test whether the complex sector can remain fully intact as an embedded sector within a larger commutative scalar algebra of minimal dimension, while allowing an additional stable analytic sector.

The complex structure of quantum mechanics was recognized early as a fundamental property [1]. The formal structure of complex Hilbert spaces was systematically developed by von Neumann [2]. Extensions of the complex scalar structure of quantum mechanics by hypercomplex algebras have been investigated in various contexts. Quaternionic extensions of quantum mechanics were studied, for example, by Adler [3].

Bicomplex and multicomplex algebras have already been extensively investigated in mathematics [4–7], and bicomplex Hilbert spaces as well as bicomplex quantum-mechanical structures have been studied in several works [8–11], but their role as possible extensions of the scalar structure of quantum mechanics has received comparatively little attention. Bicomplex structures were analyzed mathematically in detail in [4]. Algebraic properties of bicomplex numbers were presented in [5]. Hilbert spaces over bicomplex numbers were investigated, for example, in [9–11].

In contrast to earlier approaches that develop quantum-mechanical structures directly over bicomplex scalars or bicomplex Hilbert spaces [8–11], the present work focuses on the scalar algebra itself, while keeping the standard probabilistic interpretation anchored in the embedded complex sector.

While numerous approaches to extending or interpreting quantum mechanics have been proposed, the underlying scalar algebra has largely remained untouched. Extensions typically concern additional degrees of freedom, hidden variables, alternative dynamics, or modified measurement postulates. The algebraic basis of the state space itself is rarely examined systematically.

The central question of the present work is whether the complex numbers, usually taken as the scalar algebra of standard quantum mechanics, are algebraically fundamental or whether they may instead form a distinguished embedded sector of a larger commutative and associative scalar structure. The aim is therefore not to replace the complex numbers, but to determine whether they can remain fully intact inside an algebraic extension that contains an additional nontrivial sector.

It turns out that precisely such a structure is already contained in the bicomplex algebra when it is represented in a form adapted to this question. Besides the embedded complex sector, the algebra contains a proper ideal with its own complex-like analytic structure. This ideal is not merely an algebraic complement: Taylor expansions, oscillatory exponentials, Fourier-type kernels, and Schrödinger-type differential expressions can be formulated within it. Earlier work involving the present author established the existence of such analytic structures in a bicomplex ideal [7]; the present work places them into a minimal scalar-extension framework for quantum mechanics.

A particularly striking consequence is that the formal operator structure of the ideal sector leads to ideal-valued eigenvalues. For the Schrödinger-type construction considered below, the corresponding scalar belongs to the ideal sector rather than to the ordinary real scalar sector. Thus the resulting eigenvalue is genuinely ideal-valued in the full bicomplex algebra. No independent physical interpretation of such ideal-valued eigenvalues is assumed here. Nevertheless, their occurrence shows that the additional sector carries a scalar structure that is algebraically distinct from the ordinary complex sector and raises the question of whether this distinction may ultimately have physical significance.

The extension reveals a second structural feature of comparable importance. The embedded complex sector of standard quantum mechanics is not one of the canonical idempotent ideals of the bicomplex algebra. Instead, it lies diagonally across the canonical idempotent decomposition. The same minimal scalar algebra therefore carries two distinct but simultaneously relevant decompositions: the algebraically canonical decomposition into orthogonal idempotent ideals and the decomposition that keeps the ordinary complex scalar sector explicitly visible and separates it from the additional ideal-valued sector. Whether this algebraic dual structure reflects a deeper physical organization remains an open question. The present work establishes the mathematical framework in which that question can be posed precisely.

The present work therefore formulates a minimality problem for finite-dimensional real algebras. We require an algebra that

- contains the complex plane as an embedded subalgebra,
- possesses an additional nontrivial sector parallel to the embedded complex sector,
- carries this sector as a real algebra internally isomorphic to $\mathbb{C}$,
- makes this sector stable under multiplication by the extended scalar algebra,
- remains analytically viable with respect to exponential maps, Taylor series, Fourier-type kernels, and Schrödinger-type differential expressions,
- and preserves commutativity and associativity.

The stability condition implies that the additional sector is an ideal. Since its intersection with the embedded complex subalgebra is trivial, it does not contain the unit and is therefore a proper ideal.

The last requirement in this list is essential for the present investigation. Noncommutative extensions such as quaternionic quantum mechanics [3] lead to substantial structural modifications of the operator algebra and of the spectral framework. The aim here is different: to identify the **minimally invasive commutative scalar extension** that can support an additional analytic sector while leaving the standard complex sector intact.

It is shown that these requirements force a real algebra of at least four dimensions and that the bicomplex algebra

$$\mathbb{B} = \mathbb{C} \otimes_{\mathbb{R}} \mathbb{C}$$

satisfies them in minimal form. Under the additional compatibility assumption that the ideal is a unital real algebra whose internal unit is represented by an idempotent element of the full algebra, the four-dimensional realization is unique up to isomorphism.

The resulting structure should not be understood as a complete alternative dynamics of quantum mechanics. Rather, it identifies a minimal dual scalar framework in which the ordinary complex sector coexists with an ideal-valued sector. As will be shown below, this sector supports a formal ideal-valued eigenvalue structure, expressed through eigenvalue relations whose scalar values lie in the ideal sector of the bicomplex algebra. These features motivate further investigation of the possible significance of the resulting structure.

## 2 The Bicomplex Scalar Algebra

In this section, the bicomplex algebra is introduced as the candidate scalar algebra for the minimal extension considered in this work. The representation is chosen so that the idempotent element *k* and the associated additional sector, which will be shown to be an ideal, become explicit.

We consider the real algebra

$$\mathbb{B} = \mathbb{C} \otimes_{\mathbb{R}} \mathbb{C}.$$

It is four-dimensional over $\mathbb{R}$ with basis

$$\{1, i, j, k\}.$$

The multiplication rules are

$$i^2 = -1, \qquad k^2 = k, \qquad j = ik.$$

From this it follows that

$$j^2 = (ik)(ik) = i(ki)k = i(ik)k = i^2k^2 = (-1)k = -k.$$

Furthermore, we have

$$kj = k(ik) = (ki)k = (ik)k = jk = j.$$

The following multiplication table from [7] provides an overview:

*Table 1. Multiplication table of the bicomplex algebra in the basis adapted to the idempotent element k.*

| $a \cdot b$ | *1* | *i* | *j* | *k* |
|---|---|---|---|---|
| *1* | *1* | *i* | *j* | *k* |
| *i* | *i* | *-1* | *-k* | *j* |
| *j* | *j* | *-k* | *-k* | *j* |
| *k* | *k* | *j* | *j* | *k* |

This representation is isomorphic to the standard bicomplex algebra with two commuting imaginary units. The present basis is chosen because the non-real idempotent element $k$ makes the ideal structure explicit. Although $k$ is not a real scalar, it is not an imaginary unit in the usual sense. The element $j$ satisfies $j^2 = -k$ and therefore plays the role of an imaginary generator within the ideal sector. (In the standard bicomplex notation with two commuting imaginary units $i$ and $h$, $i^2 = h^2 = -1$, the present idempotent may be identified with $k = (1 + ih)/2$, and $j = ik = (i - h)/2$.)

The algebra is commutative and associative. The relation

$$k^2 = k$$

shows that $k$ is idempotent.

# 3 Ideal and Idempotent Structure

The essential structural feature of the bicomplex scalar algebra is the coexistence of an embedded complex sector and an additional sector with complex-like structure. This distinction is central for separating the ordinary complex scalar structure from the additional ideal-valued sector considered in this work. We now show that, in the bicomplex realization, this additional sector is in fact a proper ideal.

The algebra decomposes as a direct real vector space:

$$\mathbb{B} = C_1 \oplus J_1,$$

with

$$C_1 = \mathrm{Span}_{\mathbb{R}}\{1, i\}, \qquad J_1 = \mathrm{Span}_{\mathbb{R}}\{k, j\}.$$

For every $x \in J_1$ and every $a \in \mathbb{B}$ it holds that

$$ax \in J_1.$$

Since $1 \notin J_1$, the subspace $J_1$ is a proper ideal.

## 3.1 Idempotent Ideal Decomposition

From the idempotency relation

$$k^2 = k$$

it follows immediately that the elements $k$ and $1 - k$ form orthogonal idempotents.

Indeed,

$$(1-k)^2 = 1-k$$

and

$$k(1-k) = 0.$$

These idempotents induce the ideal decomposition

$$\mathbb{B} = k\mathbb{B} \oplus (1-k)\mathbb{B}.$$

This decomposition should be distinguished from the vector space decomposition

$$\mathbb{B} = C_1 \oplus J_1,$$

where $C_1 = \mathrm{Span}_{\mathbb{R}}\{1, i\}$ and $J_1 = \mathrm{Span}_{\mathbb{R}}\{k, j\}$.

The ideal $J_1$ coincides with $k\mathbb{B}$, whereas $C_1$ is the embedded complex subalgebra and is not identical with $(1-k)\mathbb{B}$. Thus, the decomposition $C_1 \oplus J_1$ must be distinguished from the idempotent ideal decomposition $k\mathbb{B} \oplus (1-k)\mathbb{B}$. The relation between these two decompositions is discussed in the next subsection.

## 3.2 Relation between the Embedded-Complex and Idempotent Decompositions

The preceding discussion shows that the bicomplex algebra admits two different decompositions. From the internal idempotent structure, one obtains the canonical ideal decomposition

$$\mathbb{B} = k\mathbb{B} \oplus (1-k)\mathbb{B}.$$

Since

$$J_1 = k\mathbb{B},$$

this may also be written as

$$\mathbb{B} = J_1 \oplus (1-k)\mathbb{B}.$$

This is the natural decomposition from the purely algebraic point of view, because both summands are ideals generated by orthogonal idempotents. In this description, however, the embedded complex sector $C_1 = \mathrm{Span}_{\mathbb{R}}\{1, i\}$ is not one of the two ideal components. Rather, $C_1$ lies diagonally across the idempotent decomposition. Indeed, for each $c \in C_1$ one has

$$c = kc + (1-k)c,$$

so that an ordinary complex scalar is split into both idempotent ideals.

For the purposes of the present work, the starting point is standard quantum mechanics. Its scalar structure is the embedded complex algebra $C_1$. We therefore use the decomposition

$$\mathbb{B} = C_1 \oplus J_1$$

as the primary decomposition. It separates the ordinary complex scalar sector of quantum mechanics from the additional proper ideal sector $J_1$. This decomposition is not the canonical decomposition into idempotent ideals, since $C_1$ is a subalgebra but not an ideal. Nevertheless, it is the decomposition adapted to the question asked here: whether the scalar algebra of quantum mechanics can be extended in a minimally invasive way while keeping the standard complex sector explicitly visible.

Relative to the decomposition

$$\mathbb{B} = C_1 \oplus J_1,$$

the complementary idempotent ideal $(1-k)\mathbb{B}$ is not aligned with $C_1$. Instead, it lies obliquely across the complex and ideal sectors. It is spanned over $\mathbb{R}$ by $1-k$ and $i-j$, since

$$i(1-k) = i - j.$$

Both generators contain a complex-sector component and an ideal-sector component:

$$1 - k = 1 + (-k),$$

$$i - j = i + (-j).$$

If $\iota: C_1 \to J_1$ denotes the natural real-algebra identification defined by

$$\iota(1) = k, \qquad \iota(i) = j,$$

then

$$(1-k)\mathbb{B} = \{c - \iota(c) \mid c \in C_1\}.$$

Thus, relative to the decomposition $C_1 \oplus J_1$, the complementary idempotent ideal is the anti-diagonal graph of the embedded complex sector.

This observation clarifies the role of the two decompositions. The idempotent decomposition $k\mathbb{B} \oplus (1-k)\mathbb{B}$ is mathematically canonical and structurally correct, but it does not by itself single out the ordinary complex scalar algebra of standard quantum mechanics. Conversely, the decomposition $C_1 \oplus J_1$ is not the canonical ideal decomposition, but it is adapted to the physical starting point of the present work. It keeps the standard complex scalar sector visible and separates it from the additional ideal-valued sector. The coexistence of these two decompositions shows that the bicomplex scalar structure used here is not merely an arbitrary realization of a possible scalar-algebra extension, but is algebraically distinguished by its internal idempotent structure.

In this sense, the relevance of the bicomplex scalar extension for quantum mechanics becomes apparent precisely when the embedded complex sector $C_1$ is distinguished from the proper ideal sector $J_1$. Precisely this decomposition will be used in the following sections in order to retain the standard Born-rule interpretation in the embedded complex sector while treating $J_1$ as the additional ideal-valued sector.

# 4 Complex-like Analytic Structure of the Ideal Sector

The preceding section identified $J_1$ as the proper ideal sector distinguished from the embedded complex scalar sector $C_1$. We now show that this ideal is not merely an algebraic subspace, but supports a complex-like analytic structure that can be used to formulate phase factors, Fourier-type kernels, and Schrödinger-type differential expressions.

The existence of a complex-isomorphic ideal within the bicomplex algebra was already investigated in [7], where it was shown that exponential functions as well as Schrödinger-like equations can be defined in this ideal.

We consider the exponential generated by $j$ in the full unital algebra $\mathbb{B}$ (Taylor expansion of $e^{jx}$):

$$e^{jx} = \sum_{n=0}^{\infty} \frac{(jx)^n}{n!}.$$

First, we compute the powers of $j$:

$$j^1 = j, \quad j^2 = -k, \quad j^3 = j^2 j = (-k)j = -j, \quad j^4 = j^2 j^2 = (-k)(-k) = k.$$

Thus, for *n ≥ 1*, the even powers satisfy

$$(jx)^{2n} = k(-1)^n x^{2n},$$

whereas the zeroth power is

$$j^0 = 1.$$

For the odd powers, *n ≥ 0*, one has

$$(jx)^{2n+1} = j(-1)^n x^{2n+1}.$$

The series therefore splits into

$$e^{jx} = 1 + k \sum_{n=1}^{\infty} (-1)^n \frac{x^{2n}}{(2n)!} + j \sum_{n=0}^{\infty} (-1)^n \frac{x^{2n+1}}{(2n+1)!}.$$

Using

$$\sum_{n=1}^{\infty} (-1)^n \frac{x^{2n}}{(2n)!} = \cos x - 1$$

and

$$\sum_{n=0}^{\infty} (-1)^n \frac{x^{2n+1}}{(2n+1)!} = \sin x,$$

one obtains

$$e^{jx} = (1 - \mathrm{k}) + k \cos x + j \sin x.$$

Within the ideal sector $k\mathbb{B}$, the idempotent $k$ acts as the internal identity element. Projecting the full exponential onto this ideal gives

$$ke^{jx} = k\cos x + j\sin x.$$

It is in this ideal-sector sense that the oscillatory structure used below is understood.

Using this ideal-sector exponential representation, one obtains in the ideal sector an oscillatory behavior similar to that of ordinary complex plane waves.

This makes it possible to construct formal Fourier-type integrals (with $\hbar$=1)

$$\psi_J(x) = \int \tilde{\psi}(p)\, k\, e^{j(px)}\, dp,$$

which mirror the standard Fourier representation of complex wave functions.

The existence of such oscillatory kernels shows that the ideal sector supports a parallel analytic phase structure rather than a merely formal algebraic extension. This property is essential for interpreting $J_1$ as a structurally distinguished ideal-valued sector in the following sections.

# 5 Minimality of the Bicomplex Scalar Structure

The central structural question is now whether the requirements formulated above single out the bicomplex algebra as the minimal scalar structure. More precisely, we ask whether a finite-dimensional, real, commutative and associative algebra can contain the ordinary complex scalar sector while also supporting an additional stable sector with complex-like analytic structure, and what the smallest possible dimension of such an algebra must be.

The minimality statement is to be understood in both a dimensional and structural sense. We therefore consider a finite-dimensional real, commutative and associative algebra $A$ with unit element satisfying the following conditions:

1. $A$ contains an embedded complex subalgebra $C \cong \mathbb{C}$,
2. $A$ possesses an additional nontrivial real sector $S$ with $S \cap C = \{0\}$,
3. $S \cong \mathbb{C}$ as a real algebra,
4. $S$ is stable under multiplication by elements of $A$,
5. the sector $S$ carries a complex-like exponential structure.

Condition 4 means that

$$aS \subset S \qquad \text{for all } a \in A.$$

Hence $S$ is an ideal of $A$. Since $S \cap C = \{0\}$ and $1 \in C$, one has $1 \notin S$; therefore $S$ is a proper ideal of $A$. In the following, we denote this ideal by $J$.

## 5.1 Impossibility of Three-Dimensional Solutions

The complex subalgebra $C$ is two-dimensional over $\mathbb{R}$. The additional sector $J$, which is required to be isomorphic to $\mathbb{C}$ as a real algebra, is also two-dimensional. Because Condition 2 requires the two sectors to have trivial intersection, their dimensions add directly. Therefore,

$$\dim_{\mathbb{R}} A \geq 4.$$

A three-dimensional solution is therefore excluded. The lower bound follows directly from requiring the additional complex-isomorphic sector to be distinct from the embedded complex subalgebra.

## 5.2 Necessity of Zero Divisors

A further structural observation concerns the existence of zero divisors.

In the present finite-dimensional setting, the existence of a nontrivial proper ideal forces the presence of zero divisors. To see this, suppose that $A$ had no zero divisors.

Multiplication by any nonzero element would then be an injective real-linear endomorphism. Because the algebra is finite-dimensional, this map would be surjective, so some element would be mapped to the unit. Hence every nonzero element would be invertible and the algebra would be a division algebra. A division algebra has no nontrivial proper ideals. Consequently, in the present finite-dimensional setting, the existence of a nontrivial proper ideal forces the presence of zero divisors.

The bicomplex algebra possesses such zero divisors explicitly. In particular,

$$k(1-k)=0,$$

although neither $k$ nor $1-k$ is equal to zero.

This property is not a defect but, in the present finite-dimensional unital setting, a structural necessity: without zero divisors no nontrivial proper ideal can exist. The requirement of a stable complex-isomorphic sector (which becomes an ideal) therefore automatically implies abandoning the division algebra property.

## 5.3 Role of Commutativity

The requirement of commutativity further restricts the structure.

Noncommutative division algebras such as the quaternions are four-dimensional and possess no nontrivial proper ideals. Clifford algebras are generally not commutative.

The simultaneous requirement of

- commutativity,
- associativity,
- a complex subalgebra,
- a stable complex-isomorphic sector, which becomes an ideal,

drastically reduces the space of admissible structures.

The bicomplex algebra satisfies these conditions in minimal dimension.

## 5.4 Minimality and Four-Dimensional Realization

In summary, the following holds:

- Two dimensions are necessary for the complex subalgebra.
- Two further dimensions are necessary for a stable complex-isomorphic sector with trivial intersection.
- Zero divisors are unavoidable.
- Commutativity excludes numerous other hypercomplex structures.

Thus, the bicomplex algebra is not merely a possible extension but, under the stated conditions, a realization of the lowest possible dimension.

In this precise sense it is a minimal commutative and associative extension of the complex scalar algebra.

The preceding minimality argument can be stated formally as follows:

Let $A$ be a real, commutative and associative algebra with the following properties:

1. $A$ contains a subalgebra $C \cong \mathbb{C}$,
2. $A$ contains an additional nontrivial real sector $S$ with $S \cap C = \{0\}$,
3. $S \cong \mathbb{C}$ as a real algebra,
4. $aS \subset S$ for all $a \in A$,
5. analytic operations such as exponential functions, Taylor series, and Fourier transformations can be defined within $S$.

Then $S$ is a proper ideal, denoted by $J$, and it necessarily follows that

$$\dim_{\mathbb{R}} A \geq 4.$$

The bicomplex algebra

$$\mathbb{B} = \mathbb{C} \otimes_{\mathbb{R}} \mathbb{C}$$

realizes this structure in exactly four real dimensions.

Thus, the bicomplex algebra provides a concrete realization of the required structure in minimal real dimension.

## 5.5 Uniqueness under a Compatibility Assumption

We now show that, under an additional compatibility assumption on the stable sector $J$ obtained in this way, the four-dimensional realization is unique up to isomorphism. More precisely, if the ideal is a unital real algebra isomorphic to $\mathbb{C}$, with its internal unit represented by an idempotent element of $A$, then every such four-dimensional algebra is isomorphic to the bicomplex algebra.

*Proof.* Let $A$ be a four-dimensional real, commutative and associative algebra with unit element that contains a subalgebra

$$C = \mathrm{Span}_{\mathbb{R}}\{1, i\} \cong \mathbb{C}, \qquad i^2 = -1$$

and a proper ideal $I \cong \mathbb{C}$ such that $C \cap I = \{0\}$. We assume in addition that $I$ is a unital real algebra isomorphic to $\mathbb{C}$, with its internal unit represented by an idempotent element $\eta \in I \subset A$.

Because

$$\dim_{\mathbb{R}} A = 4, \qquad \dim_{\mathbb{R}} C = 2, \qquad \dim_{\mathbb{R}} I = 2$$

it follows that

$$A = C \oplus I.$$

By this assumption, there exists a basis $\{\eta, \kappa\}$ of $I$ such that $\eta$ is the internal unit of $I$ and

$$\eta^2 = \eta, \quad \eta\kappa = \kappa, \quad \kappa^2 = -\eta.$$

Because $I$ is an ideal, the elements $i\eta$ and $i\kappa$ lie in $I$. Hence there exist $p, q, r, s \in \mathbb{R}$ such that

$$i\eta = p\eta + q\kappa, \quad i\kappa = r\eta + s\kappa.$$

From

$$i\kappa = i(\eta\kappa) = (i\eta)\kappa$$

one obtains

$$i\kappa = (p\eta + q\kappa)\kappa = p\kappa - q\eta,$$

which implies

$$r = -q, \quad s = p.$$

Furthermore,

$$-\eta = i^2\eta = i(i\eta) = i(p\eta + q\kappa) = p(i\eta) + q(i\kappa).$$

Substituting yields

$$-\eta = p(p\eta + q\kappa) + q(-q\eta + p\kappa) = (p^2 - q^2)\eta + 2pq\,\kappa.$$

Hence

$$p^2 - q^2 = -1, \qquad 2pq = 0.$$

Over $\mathbb{R}$ this implies

$$p = 0, \qquad q = \pm 1.$$

Thus

$$i\eta = \pm\kappa, \quad i\kappa = \mp\eta.$$

By replacing $\kappa$ with $-\kappa$ one may assume without loss of generality

$$i\eta = \kappa, \quad i\kappa = -\eta.$$

These are exactly the relations of the bicomplex algebra in the idempotent basis $\{1, i, \eta, \kappa\}$, with $\eta$ corresponding to *k* and $\kappa$ corresponding to *j*. Therefore

$$A \cong \mathbb{C} \otimes_{\mathbb{R}} \mathbb{C}.$$

Thus, the bicomplex algebra is not introduced here as an arbitrary hypercomplex choice. It appears as the minimal commutative and associative scalar extension satisfying the stated requirements: an embedded complex sector, a proper complex-like ideal, analytic viability, and an idempotent internal unit for the ideal. This is the sense in which the bicomplex structure is minimal in the present work.

# 6 Comparison with Other Hypercomplex Structures

The question of extending the complex scalar algebra of quantum mechanics is not new. Quaternionic formulations have been studied, for example, by Adler and Emch [3,12], while bicomplex and multicomplex algebras have been developed in the mathematical literature [4–6]. In order to place the present bicomplex scalar structure in context, this section briefly compares it with several familiar algebraic alternatives.

## 6.1 Quaternionic Extensions

One of the best-known alternatives is quaternionic quantum mechanics.

The quaternions form a four-dimensional real division algebra with standard basis $\{1, i, j, k\}$ and noncommutative multiplication. In a quaternionic formulation of quantum mechanics, states are defined over a noncommutative scalar domain, which has substantial structural consequences for operator products and spectral theory.

In particular, the order of scalar multiplications is no longer irrelevant. This leads to profound changes in the definition of adjoint operators as well as in the structure of unitary transformations. The noncommutativity directly affects the physical interpretation.

The bicomplex scalar structure considered here differs in this essential respect. It remains commutative and associative and therefore avoids the corresponding restructuring of scalar order, operator products, and spectral interpretation. Its purpose is to identify a minimal commutative scalar structure compatible with an additional ideal-valued analytic sector.

## 6.2 Clifford Algebras

Clifford algebras play a central role in the description of spinors and relativistic theories. They are, however, generally not commutative and primarily serve to encode geometric structures.

In contrast, the bicomplex algebra is not a geometric extension of the state space but an extension of the scalar base itself. It introduces no new geometric degrees of freedom but supplements the complex structure with a proper ideal that remains analytically viable.

## 6.3 Dual Numbers and Nilpotent Extensions

Another class of possible extensions consists of algebras containing nilpotent elements, such as the dual numbers with

$$\varepsilon^2 = 0.$$

Such structures allow interesting algebraic constructions and possess exponential maps, but they do not provide the complex-like oscillatory exponential structure required here.

In particular, the power series expansion of

$$e^{\varepsilon x}$$

produces linear rather than trigonometric terms. Consequently, the cyclic structure characteristic of interference phenomena is absent.

By contrast, the ideal $J_1$ of the bicomplex algebra satisfies the relation

$$j^2 = -k,$$

which leads to a complex-like exponential expansion. This property distinguishes the bicomplex structure from purely nilpotent extensions.

## 6.4 Minimality in Comparison

The combination of

- commutativity,
- associativity,
- the existence of a proper ideal,
- a complex-isomorphic structure of the ideal,
- and analytic viability

is highly restrictive.

Quaternions do not satisfy the commutativity condition. Clifford algebras are generally not commutative and serve a different geometric purpose. Purely nilpotent sectors, such as the dual-number sector, do not possess the complex-like oscillatory exponential structure required here.

Under the stated criteria, the bicomplex algebra is therefore structurally distinguished. It is not selected as the most general hypercomplex extension, but as a minimal commutative and associative scalar structure that retains the ordinary complex sector while adding a proper complex-like ideal. This is precisely the sense in which the bicomplex extension is minimally invasive in the present work.

# 7 Bicomplex Module Structure and Born Rule

The extended state space is defined as

$$\mathcal{M} = \mathcal{H} \otimes_{\mathbb{C}} \mathbb{B}.$$

Here $\mathbb{B}$ is regarded as a $\mathbb{C}$-algebra via the canonical embedding

$$\mathbb{C} \cong C_1 \subset \mathbb{B},$$

so that the tensor product is well defined.

## 7.1 Module Structure of the Extended State Space

The extended state space $\mathcal{M}$ is constructed as a module over the bicomplex algebra $\mathbb{B}$. A module generalizes the concept of a vector space: while the scalars of a vector space form a field (all nonzero elements are invertible), the scalars of a module may come from an algebra—even if that algebra contains zero divisors.

Let $\mathcal{A}$ be a commutative, associative $\mathbb{R}$-algebra with unit element $1_{\mathcal{A}}$. An $\mathcal{A}$-module is an abelian group $M$ together with a scalar multiplication

$$\mathcal{A} \times M \to M, \qquad (a, \psi) \mapsto a\psi,$$

such that for all $a, b \in \mathcal{A}$ and $\psi, \phi \in M$ the following relations hold:

1. $(a + b)\psi = a\psi + b\psi$
2. $a(\psi + \phi) = a\psi + a\phi$
3. $(ab)\psi = a(b\psi)$
4. $1_{\mathcal{A}}\psi = \psi$.

In the present work we take

$$\mathcal{A} = \mathbb{B} = \mathbb{C} \otimes_{\mathbb{R}} \mathbb{C}.$$

The state space is

$$\mathcal{M} = \mathcal{H} \otimes_{\mathbb{C}} \mathbb{B},$$

where $\mathcal{H}$ is a complex Hilbert space. The tensor product is taken over $\mathbb{C}$, using the canonical embedding $\mathbb{C} \hookrightarrow \mathbb{B}$.

The space $\mathcal{M}$ carries two compatible module structures:

1. As a $\mathbb{C}$-module:

$$\lambda(\psi \otimes b) = (\lambda\psi) \otimes b = \psi \otimes (\lambda b), \qquad \lambda \in \mathbb{C},$$

2. As a $\mathbb{B}$-module:

$$a(\psi \otimes b) = \psi \otimes (ab), \qquad a \in \mathbb{B}.$$

*Proof.* The module axioms follow directly from the properties of the tensor product and the multiplication in $\mathbb{B}$. Compatibility follows from the embedding $\mathbb{C} \subset \mathbb{B}$.

The probabilistic interpretation considered in the present work remains anchored in the complex sector:

Every state decomposes uniquely as

$$\Psi = \Psi_C + \Psi_J.$$

Here, this decomposition refers to the vector-space decomposition

$$\mathbb{B} = C_1 \oplus J_1,$$

introduced above. As discussed earlier it should be distinguished from the idempotent ideal decomposition

$$\mathbb{B} = k\mathbb{B} \oplus (1-k)\mathbb{B}.$$

## 7.2 Born Rule in the Embedded Complex Sector

For the purpose of recovering the standard Born rule, only the embedded complex component is used. This gives the complex-valued projection formula

$$\langle \Psi \mid \Phi \rangle_{\text{phys}} = \langle \Psi_C \mid \Phi_C \rangle_{\mathcal{H}}.$$

In this restricted sense, the usual Born rule remains unchanged. No separate probabilistic interpretation of the ideal component is introduced here.

## 7.3 Canonical Decomposition of the Extended State Space

The previous discussion shows that the scalar extension need not modify the standard probabilistic interpretation as long as this interpretation is restricted to the complex sector. However, a structural question arises concerning the algebraic form of the extended state space.

The following statement shows that the bicomplex state space possesses a particularly simple structure.

Let $\mathcal{H}$ be a complex Hilbert space and $\mathcal{M} = \mathcal{H} \otimes_{\mathbb{C}} \mathbb{B}$ the state space over the bicomplex algebra $\mathbb{B} = \mathbb{C} \otimes_{\mathbb{R}} \mathbb{C}$. Then $\mathcal{M}$ admits a canonical decomposition

$$\mathcal{M} \cong \mathcal{H} \oplus \mathcal{H}.$$

*Proof.* The bicomplex algebra contains an idempotent element satisfying

$$k^2 = k.$$

This implies

$$(1 - k)^2 = 1 - k$$

and

$$k(1 - k) = 0.$$

The elements $k$ and $1 - k$ are orthogonal idempotents and induce complementary projection maps by multiplication. Consequently, the algebra decomposes as the direct sum of two ideals

$$\mathbb{B} = k\mathbb{B} \oplus (1 - k)\mathbb{B}.$$

Substituting this decomposition into the state space yields

$$\mathcal{M} = \mathcal{H} \otimes_{\mathbb{C}} \mathbb{B} = \mathcal{H} \otimes_{\mathbb{C}} (k\mathbb{B} \oplus (1-k)\mathbb{B}).$$

The tensor product distributes over direct sums, hence

$$\mathcal{M} \cong (\mathcal{H} \otimes_{\mathbb{C}} k\mathbb{B}) \oplus (\mathcal{H} \otimes_{\mathbb{C}} (1-k)\mathbb{B}).$$

Since both $k\mathbb{B}$ and $(1-k)\mathbb{B}$ are isomorphic to $\mathbb{C}$ as $\mathbb{C}$-modules, it follows that

$$\mathcal{H} \otimes_{\mathbb{C}} k\mathbb{B} \cong \mathcal{H}, \qquad \mathcal{H} \otimes_{\mathbb{C}} (1-k)\mathbb{B} \cong \mathcal{H}.$$

Thus, one obtains

$$\mathcal{M} \cong \mathcal{H} \oplus \mathcal{H}.$$

The bicomplex extension of the state space therefore does not lead to an unstructured enlargement of the Hilbert space. Rather, with the decomposition induced by $\mathbb{B} = k\mathbb{B} \oplus (1 - k)\mathbb{B}$, the canonical decomposition $M \cong \mathcal{H} \oplus \mathcal{H}$ corresponds to the two idempotent ideal components $\mathcal{H} \otimes_{\mathbb{C}} k\mathbb{B} = \mathcal{H} \otimes_{\mathbb{C}} J_1$ and $\mathcal{H} \otimes_{\mathbb{C}} (1 - k)\mathbb{B}$, respectively. In particular, the second summand is not $\mathcal{H} \otimes_{\mathbb{C}} C_1$. This canonical decomposition is therefore distinct from the vector-space decomposition used above to separate the embedded complex scalar sector from the ideal sector.

A purely ideal-sector amplitude does not yield an ordinary real-valued probability within the standard Born-rule construction used here. Algebraic products formed entirely within the ideal sector remain ideal-valued and contain the idempotent factor $k$. In the present work, such quantities are not assigned a separate probabilistic interpretation.

# 8 Formal Operator Structure in the Ideal Sector

After the Born-rule interpretation has been restricted to the embedded complex sector, it remains to examine whether the ideal sector carries a formally analogous operator structure. This question is purely algebraic: no independent spectral or measurement interpretation of ideal-valued quantities is assumed.

Throughout this section we use natural units $\hbar = 1$. For

$$\psi_J = ke^{j(px-Et)}$$

one obtains

$$\partial_t \psi_J = -jE\psi_J.$$

With

$$\hat{H}_J = j\frac{\partial}{\partial t}$$

it follows that

$$\hat{H}_J\psi_J = j(-jE)\psi_J = -(j^2)E\psi_J = kE\psi_J.$$

The choice of the operator

$$\hat{H}_J = j\frac{\partial}{\partial t}$$

corresponds to the formal analogy with the complex Schrödinger structure, where $j$ plays the role of the imaginary unit in the ideal sector.

The ideal sector therefore possesses a formal operator eigenvalue structure. The corresponding eigenvalue relation is ideal-valued. Within $J_1$, the idempotent $k$ acts as the internal identity. Nevertheless, for $E \neq 0$ , $kE$ and $E$ are distinct elements of the full bicomplex algebra: $E$ belongs to the embedded real scalar sector, whereas $kE$ belongs to the ideal sector $J_1$. Thus, the operator relation carries a genuinely ideal-valued scalar when viewed in the full bicomplex algebra. Whether this algebraic distinction is associated with distinct physical properties remains an open question. No spectral interpretation beyond this formal algebraic relation is assumed here.

# 9 Algebraic Phase Relation and Anti-Diagonal Sector

The multiplication rules of the bicomplex algebra imply an algebraic phase relation between the embedded complex sector and the ideal sector.

One finds

$$e^{ix}ke^{jy} = (\cos x + i \sin x)\, k(\cos y + j \sin y).$$

After expanding the expression and using the relations

$$ik = j$$

and

$$ij = -k,$$

one obtains

$$k(\cos x \cos y - \sin x \sin y) + j(\cos x \sin y + \sin x \cos y).$$

Using the addition formulas yields

$$ke^{j(x+y)}.$$

The multiplicative structure therefore implies an algebraic phase relation between the two sectors. This relation is purely algebraic and no physical interpretation of it is assumed in the present work.

A related observation concerns phase-matched elements in the two sectors. Let

$$\iota: C_1 \to J_1$$

be the natural real-algebra identification defined by

$$\iota(1) = k, \qquad \iota(i) = j.$$

For a complex phase factor $e^{i\theta}$, its ideal-sector image is

$$\iota\left(e^{i\theta}\right) = k\cos\theta + j\sin\theta.$$

Equivalently,

$$\iota\left(e^{i\theta}\right) = ke^{j\theta}.$$

The anti-diagonal combination

$$\Omega_\theta = e^{i\theta} - \iota\left(e^{i\theta}\right)$$

therefore satisfies

$$\Omega_\theta = e^{i\theta} - ke^{j\theta} = (1-k)\cos\theta + (i-j)\sin\theta.$$

Hence

$$\Omega_\theta \in (1-k)\mathbb{B}.$$

Thus, the complementary idempotent ideal $(1-k)\mathbb{B}$ can be represented as the anti-diagonal sector generated by phase-matched complex and ideal components. This observation is again purely algebraic. It does not introduce a new physical superposition principle or a separate measurement interpretation. Nevertheless, from a purely algebraic point of view, $\Omega_\theta$ represents a structurally interesting algebraic object within the extended scalar structure.

# 10 Entangled States as an Algebraic Test Case

Entangled states provide a useful algebraic test case for the extended module framework. The purpose of this section is not to modify the standard tensor-product structure or to propose a new interpretation of nonlocal correlations, but to examine how the bicomplex scalar extension represents separable and nonseparable states algebraically.

For two subsystems the extended state space becomes

$$\mathcal{M}_{AB} = (\mathcal{H}_A \otimes \mathcal{H}_B) \otimes_{\mathbb{C}} \mathbb{B}.$$

This space contains the ordinary complex tensor product Hilbert space as a substructure and, in addition, an ideal-valued component. Every state can be decomposed as

$$\Psi = \Psi_C + \Psi_J,$$

where both components are defined over the same tensor-product structure.

## 10.1 Separability and Ideal Structure

A separable complex state has the form

$$|\Phi\rangle = |\phi\rangle_A \otimes |\chi\rangle_B.$$

As an illustrative bicomplex extension that keeps the original complex state and adds an ideal-sector copy, one may consider multiplication by the factor $1 + k$:

$$|\Phi\rangle_{\mathbb{B}} = (1 + k)\,|\phi\rangle_A \otimes |\chi\rangle_B.$$

Equivalently,

$$|\Phi\rangle_{\mathbb{B}} = |\phi\rangle_A \otimes |\chi\rangle_B + k|\phi\rangle_A \otimes |\chi\rangle_B.$$

Both the complex component and the ideal component factorize. Thus, the bicomplex extension does not by itself generate entanglement from a separable state. The product structure remains preserved.

This shows that the algebraic extension respects separability in this construction: separable states remain separable when the ideal component is introduced by a common scalar factor.

## 10.2 The Singlet State

The singlet state

$$|\Psi^-\rangle = \frac{1}{\sqrt{2}}(|\uparrow\rangle_A|\downarrow\rangle_B - |\downarrow\rangle_A|\uparrow\rangle_B)$$

is not separable. Its anticorrelation is already present in the ordinary complex tensor-product structure.

Using the same illustrative extension, one obtains

$$|\Psi^-\rangle_{\mathbb{B}} = (1+k)\,|\Psi^-\rangle.$$

This gives

$$|\Psi^-\rangle_{\mathbb{B}} = |\Psi^-\rangle + k|\Psi^-\rangle.$$

The first term lies in the complex sector, whereas the second term lies in the ideal sector.

Explicitly,

$$k|\Psi^-\rangle = \frac{k}{\sqrt{2}}(|\uparrow\rangle_A|\downarrow\rangle_B - |\downarrow\rangle_A|\uparrow\rangle_B).$$

Since $k$ acts here as a common scalar factor in the extended module, it does not change the tensor structure of the singlet state. The antisymmetric nonseparable structure is therefore preserved in the ideal component.

Applying the projector to subsystem $A$ yields

$$\Pi_\uparrow^{(A)}|\Psi^-\rangle_{\mathbb{B}} = \frac{1+k}{\sqrt{2}}\,|\uparrow\rangle_A|\downarrow\rangle_B,$$

and analogously for the opposite projection.

Thus, in this algebraic representation, the same antisymmetric correlation pattern is mirrored in both the complex component and the ideal component.

## 10.3 Role of Idempotency

The idempotency relation

$$k^2 = k$$

implies that multiplication by $k$ defines a projector onto the ideal $J_1 = k\mathbb{B}$. Moreover, $k$ acts as the internal identity on $J_1$: for every $x \in J_1$, $kx = x$.

The stability of $J_1$ under multiplication by arbitrary elements of $\mathbb{B}$, by contrast, follows from its ideal property and not from idempotency alone.

In the illustrative extension used above, multiplication by $1 + k$ produces a state with two algebraic components,

$$(1 + k) \mid \Psi^{-}\rangle = \mid \Psi^{-}\rangle + k \mid \Psi^{-}\rangle,$$

without changing the internal tensor structure of the singlet state.

The purpose of this example is not to introduce a new physical normalization or a new interpretation of measurement. It only illustrates that the same algebraic correlation structure can be represented simultaneously in the complex component and in the ideal component. Relative to the original complex Hilbert-space structure, the additional ideal-sector term $k \mid \Psi^{-}\rangle$ is not obtained by an ordinary complex rescaling.

## 10.4 Correlation Structure in the Ideal Sector

The bicomplex extension does not generate the anticorrelation of the singlet state. The anticorrelation is already fully present in the ordinary complex state vector. However, the extended module provides an algebraic representation in which the same nonseparable correlation structure is also carried by the ideal component. Thus, entanglement constitutes a natural structural contact point between the standard tensor-product framework and the ideal sector of the bicomplex extension.

This does not amount to an alternative explanation of nonlocality or to a modification of the standard quantum-mechanical predictions. The result is only that the extended scalar algebra allows the algebraic form of an entangled state to be represented simultaneously in the embedded complex sector and in the ideal sector. Whether this additional ideal representation has significance beyond the algebraic level is left open.

# 11 Summary and Discussion

The present work has investigated whether the complex scalar algebra of quantum mechanics can remain fully embedded within a minimal commutative extension containing an additional complex-like ideal sector. The analysis shows that this requirement does not lead to an arbitrary hypercomplex enlargement, but to a highly constrained algebraic structure distinguished by its minimality, ideal structure, and idempotent decomposition.

It was shown that the simultaneous requirement of

- a complex subalgebra,
- an additional nontrivial sector parallel to it,
- a complex-isomorphic structure of this sector,
- stability under the extended scalar multiplication,
- analytic viability,
- commutativity and associativity,

forces the existence of a real algebra of at least four dimensions. The stability condition identifies the additional sector as an ideal, while its trivial intersection with the embedded complex subalgebra ensures that the ideal is proper.

The bicomplex algebra identified here is therefore not introduced as an arbitrary hypercomplex structure, but as a minimal scalar framework satisfying the stated structural requirements. It contains the ordinary complex sector together with an ideal-valued sector, while preserving commutativity, associativity, analytic viability, and the standard probabilistic interpretation in the embedded complex sector.

The role of the ideal-valued sector $J_1$ remains formal and algebraic in the present work. Nevertheless, $J_1$ is not merely an auxiliary component: it is tied to the embedded complex sector $C_1$ through the idempotent structure of $\mathbb{B}$ and through the anti-diagonal relation involving $(1-k)\mathbb{B}$. In this sense, $J_1$ appears as a structurally distinguished part of the minimal bicomplex scalar framework. This motivates further mathematical study, without amounting to an independent physical interpretation or a physical existence claim.

## 11.1 Structural Particularities

The bicomplex scalar algebra combines several restrictive properties in a single finite-dimensional structure. It is commutative and associative, contains the embedded complex sector $C_1$, and possesses the proper ideal $J_1 = k\mathbb{B}$. This ideal carries a complex-like analytic structure, while the multiplication rules also generate an algebraic phase relation between the complex and ideal sectors. In addition, the ideal sector supports a formal ideal-valued eigenvalue relation in which the corresponding scalar remains ideal-valued in the full algebra.

The coexistence of these properties is structurally restrictive. Many other hypercomplex extensions fail to preserve either commutativity, the embedded complex scalar sector, or the complex-like analytic behavior required here. This is why the bicomplex algebra is not merely a convenient example, but a structurally distinguished realization of the minimal scalar framework considered in this work.

## 11.2 Phase Relation and Anti-Diagonal Structure

The algebraic phase relation derived in Section 9 is one of the characteristic consequences of the bicomplex multiplication rules. If a complex phase factor in the embedded sector is multiplied with an ideal-sector phase factor, the result remains in the ideal sector and the phases add algebraically:

$$e^{i\varphi_C}\, k e^{j\varphi_J} = k e^{j(\varphi_C+\varphi_J)}.$$

Thus, the complex phase does not act independently of the ideal phase. Rather, multiplication in the full bicomplex algebra induces an additive phase relation between the two sectors. This is a purely algebraic statement, but it shows that the embedded complex sector and the ideal-valued sector are not simply placed side by side.

A formal condition such as

$$\varphi_C + \varphi_J = c_0,$$

with $c_0$ constant, may be used as a purely formal assumption to couple the two sectors algebraically through this phase relation. Under such a constraint, changes in one phase would be compensated by changes in the other within the algebraic expression. No physical realization of this constraint, and no dynamical mechanism behind it, is assumed here.

A second, and in the present context more intrinsic, possibility is phase matching. If

$$\varphi_C = \varphi_J = \theta,$$

then the natural real-algebra identification

$$\iota: C_1 \to J_1$$

maps the ordinary complex phase to the corresponding ideal-sector phase,

$$\iota\left(e^{i\theta}\right) = ke^{j\theta}.$$

The anti-diagonal combination

$$\Omega_\theta = e^{i\theta} - \iota\left(e^{i\theta}\right)$$

therefore satisfies

$$\Omega_\theta = e^{i\theta} - ke^{j\theta} = (1-k)\cos\theta + (i-j)\sin\theta.$$

Hence

$$\Omega_\theta \in (1-k)\mathbb{B}.$$

This shows that the anti-diagonal sector is generated by phase-matched differences between the embedded complex sector and its ideal-sector image. In this sense, the anti-diagonal structure is not independent of the phase relation, but provides a second algebraic way in which the two sectors are connected.

Both observations remain formal. The additive phase law itself is an algebraic consequence of bicomplex multiplication, whereas conditions such as $\varphi_C + \varphi_J = c_0$ and $\varphi_C = \varphi_J$ are algebraic possibilities that may be considered within this structure. They do not introduce a new physical phase postulate, but they show that the ideal-valued sector is structurally tied to the ordinary complex sector through the internal multiplication and decomposition structure of $\mathbb{B}$.

## 11.3 Formal Operator Perspective

The formal operator construction considered in Section 8 shows that the ideal sector $J_1$ is closed under a Schrödinger-type algebraic operation.

In this limited sense, $J_1$ carries a formal ideal-valued eigenvalue relation of the form

$$\widehat{H}_J \psi_J = kE\, \psi_J.$$

Within $J_1$, the idempotent $k$ acts as the internal identity element, so that $kE$ plays the role of the scalar corresponding to $E$ inside the ideal sector. As an element of the full bicomplex algebra, however, $kE$ remains ideal-valued rather than real-valued. It is therefore not an ordinary real energy eigenvalue in the standard quantum-mechanical sense.

The physical meaning of states or functions with non-real, ideal-valued eigenvalues remains open and is not investigated in the present work. The operator relation should therefore be understood as a formal algebraic analogue, not as an independent spectral interpretation. The Born-rule restriction to the embedded complex sector remains unchanged, and ideal-valued products are not assigned a separate probabilistic interpretation here.

## 11.4 Entangled States as a Structural Contact Point

Entangled states provide a natural structural contact point between the standard tensor-product framework and the ideal-valued sector of the bicomplex extension. The analysis in Section 10 does not suggest that the bicomplex scalar extension generates entanglement or modifies the standard nonseparable structure. That structure is already fully present in the ordinary complex Hilbert space.

What the extended module provides is an algebraic representation in which the same nonseparable tensor structure can be accompanied by an ideal-valued component. In this sense, the ideal sector mirrors an existing quantum correlation structure without changing the standard quantum-mechanical predictions. This observation remains purely algebraic and does not amount to an alternative explanation of nonlocality or to a new dynamics.

## 11.5 Open Mathematical Questions

The present analysis suggests several mathematical questions for further work. A complete spectral theory for the specific projected-probability framework considered here remains to be developed. In particular, the status of ideal-valued eigenvalues of the form $kE$ requires clarification beyond the formal algebraic relation considered here.

A second question concerns the role of idempotent and projector-like elements in the structure of unitary transformations. Within the specific projected-probability framework considered here, the presence of nontrivial idempotents and zero divisors means that the usual complex Hilbert-space notions of normalization, adjointness, and unitarity require separate analysis.

A third question concerns the anti-diagonal sector $(1-k)\mathbb{B}$. As shown above, this sector can be represented through phase-matched differences between the embedded complex sector and its ideal-sector image. It remains open whether this anti-diagonal structure has a deeper role in the module theory or in the representation of correlations.

Finally, one may ask whether natural representations of physical symmetries exist within the extended scalar structure. These questions concern the mathematical structure of the bicomplex framework and are independent of any physical interpretation.

## 11.6 Physical Perspective

The present work does not claim to propose a modified quantum mechanics. The standard theory remains fully intact within the embedded complex sector $C_1$, and the Born rule is retained there. The ideal-valued sector $J_1$ is not assigned an independent measurement or probability interpretation in the present work.

At the same time, the ideal sector is not structurally empty. It carries a complex-like analytic structure, a formal operator relation with ideal-valued eigenvalues, an algebraic phase relation to the embedded complex sector, and a connection to the anti-diagonal idempotent sector. These features suggest that $J_1$ should be regarded as a mathematically significant sector of the minimal bicomplex scalar framework.

From the purely algebraic point of view, the canonical decomposition of $\mathbb{B}$ is the idempotent ideal decomposition $\mathbb{B} = k\mathbb{B} \oplus (1 - k)\mathbb{B}$. In this limited algebraic sense, the idempotent decomposition may be regarded as the more fundamental decomposition of the bicomplex algebra. The embedded complex sector $C_1$, on which the standard Born-rule interpretation is retained, is therefore not one of the canonical ideal components, but is positioned diagonally with respect to this decomposition. This mismatch between the algebraically canonical decomposition and the physically selected complex sector is one of the structural features that motivates further study.

Whether this algebraic structure has physical significance beyond the formal scalar extension is left open. The present article establishes the minimal bicomplex scalar framework and identifies the ideal-valued sector as a structurally distinguished part of that framework. Questions concerning physical realizability, interpretation, or empirical relevance lie beyond the scope of the present work.

# Acknowledgments

The author thanks Ulm University of Applied Sciences for its support. The author also thanks Professor Viktor Otte, emeritus professor at the University of Wuppertal, for valuable comments and stimulating discussions.

# Statements and Declarations

Funding

The author received no external funding for the research or authorship of this article.

Competing Interests

The author declares that he has no competing interests.

Data Availability

No datasets were generated or analyzed during the current study.